\documentclass[11pt]{article}
\usepackage[letterpaper,hmargin=26mm,vmargin=1in]{geometry}
\usepackage{graphicx,amssymb,amsmath,amsthm}

\usepackage{algorithm}
\usepackage{algorithmic}

\usepackage{booktabs}
\usepackage{longtable,array}
\usepackage{latexsym}
\usepackage{epic}
\usepackage{eepic}

\newtheorem{theorem}{Theorem}[section]

\newtheorem{proposition}[theorem]{Proposition}
\newtheorem{corollary}[theorem]{Corollary}

\theoremstyle{definition}
\newtheorem{definition}[theorem]{Definition}
\newtheorem{definitions}[theorem]{Definitions}
\newtheorem{example}[theorem]{Example}

\newcommand{\ignore}[1]{}

\newcommand{\ceil}[1]{\lceil #1 \rceil }

\newcommand\eat[1]{}

\begin{document}

\title{Computing voting power in easy weighted voting games}

\author{
\em Haris Aziz
\and \em Mike Paterson
}

\date{} 

\maketitle

\begin{abstract}
\emph{Weighted voting games} are ubiquitous mathematical models which are used in economics, political science, neuroscience, threshold logic, reliability theory and distributed systems. They model situations where agents with variable voting weight vote in favour of or against a decision. A coalition of agents is winning if and only if the sum of weights of the coalition exceeds or equals a specified quota. The Banzhaf index is a measure of voting power of an agent in a weighted voting game. It depends on the number of coalitions in which the agent is the difference in the coalition winning or losing. It is well known that computing Banzhaf indices in a weighted voting game is NP-hard. We give a comprehensive classification of weighted voting games which can be solved in polynomial time. Among other results, we provide a polynomial ($O(k{(\frac{n}{k})}^k)$) algorithm to compute the Banzhaf indices in weighted voting games in which the number of weight values is bounded by $k$. 
\end{abstract}

\section{Introduction}

\subsection{Motivation and Background}

Weighted voting games (WVGs) are mathematical models which are used to analyze voting bodies in which the voters have different number of votes.  In WVGs, each voter is assigned a non-negative weight and makes a vote in favour of or against a decision. The decision is made if and only if the total weight of those voting in favour of the decision is greater than or equal to some fixed quota. Since the weights of the players do not always exactly reflect how critical a player is in decision making, voting power attempts to measure the ability of a player in a WVG to determine the outcome of the vote. WVGs are also encountered in threshold logic, reliability theory, neuroscience and logical computing devices \cite{simplegames,45427}. Parhami~\cite{parhami2005yes} points out that voting has a long history in reliability systems dating back to von Neumann~\cite{vN56}. For reliability systems, the weights of a WVG can represent the significance of the components whereas the quota can represent the threshold for the overall system to fail. 
WVGs have been applied in various political and economic organizations~\cite{bilbao_2007, Aziz2007a}. Voting power is also used in joint stock companies where each shareholder gets votes in proportion to the ownership of a stock \cite{gambarelli2}. 

The Banzhaf index is considered the most suitable power index by voting power theorists \cite{RePEc:bla:polstu:v:50:y:2002:i:1:p:1-22, mainbook}. The computational complexity of computing Banzhaf indices in WVGs is well studied. Prasad and Kelly~\cite{91928} show that the problem of computing the Banzhaf values of players is \#P-complete. It is even NP-hard to identify a player with zero voting power or two players with same Banzhaf indices \cite{matsui00survey}. Klinz and Woeginger~\cite{RePEc:eee:matsoc:v:49:y:2005:i:1:p:111-116} devised the fastest exact algorithm to compute Banzhaf indices in a WVG. In the algorithm, they applied a partitioning approach that dates back to Horowitz and Sahni \cite{321823}. However the complexity of the algorithm is still $O(n^2 2^{\frac{n}{2}})$. In this paper, we restrict our analysis to exact computation of Banzhaf indices instead of examining approximate solutions.
We show that although computing Banzhaf indices of WVGs is a hard problem in general, it is easy for various classes of WVGs, e.g., for WVGs with a bounded number of weight values,  
an important sub-class of WVGs. 

\subsection{Outline}
Section \ref{sec:pre} provides the preliminary definitions of terms used in the paper. 
The outline of the paper is as following.
Section~\ref{SecClassConst} identifies WVGs in which Banzhaf indices can be computed in constant time. In Section~\ref{SecClassBounded}, we examine WVGs with a bounded number of weight values, and provide algorithms to compute the Banzhaf indices. Section~\ref{SecClassDist} examines WVGs with special weight distributions. 
Section~\ref{SecClassInteger} considers WVGs with integer weights.
We conclude with some open problems in the final section.

\section{Preliminaries}
\label{sec:pre}

We give definitions of key terms. The set of voters is $N = \{1,...,n\}$. 

\begin{definitions}
A \emph{simple voting game} is a pair $(N,v)$ with $v:2^N \rightarrow \{0,1\}$ where $v(\emptyset)=0$, $v(N)=1$ and $v(S)\leq v(T)$ whenever $S \subseteq T$. A coalition $S \subseteq N$ is \emph{winning} if $v(S)=1$ and \emph{losing} if $v(S)=0$. A simple voting game can alternatively be defined as $(N,W)$ where $W$ is the set of winning coalitions.
\end{definitions}
\begin{definition}
The simple voting game $(N,v)$ where \\
$W=\{X \subseteq N, \sum_{x \in X}{w_x} \geq q \}$ is called a \emph{weighted voting game}. A weighted voting game is denoted by $[q;w_1,w_2,... ,w_n]$ where $w_i$ is the voting weight of player $i$. Usually, $w_i \geq w_j$ if $i<j$.
\end{definition}

Generally, $\frac12 \sum_{1 \leq i \leq n}w_i \leq q \leq {\sum_{1 \leq i \leq n}w_i}$ so that there can be no two disjoint winning coalitions. Such weighted voting games are termed \emph{proper}. 

\begin{definitions}
A player $i$ is \emph{critical} in a coalition $S$ when $S \in W$ and $(S \setminus i) \notin W$. For each $i \in N$, we denote the number of coalitions in which $i$ is critical in game $v$ by ${{\eta}_{i}}(v)$. The \emph{Banzhaf index} of player $i$ in weighted voting game $v$ is $\beta_i = \frac{{{\eta}_{i}}(v)}{{\sum}_{i \in N}{{\eta}_{i}}(v)}$. The \emph{probabilistic Banzhaf index}, $\beta_i^{'}$ of player $i$ in game $v$ is ${{{\eta}_{i}}(v)}/2^{n-1}$. \emph{Coleman's power of the collectivity to act}, $A$, is defined as the ratio of the number of winning coalitions $w$ to $2^n$: $A = w/2^n $.
\end{definitions}

The problem of computing the Banzhaf indices of a WVG can be defined formally as following:

\noindent
\textbf{Name}: BI-WVG \\
\textbf{Instance}: WVG, $v = [q; w_1,... ,w_n]$\\
\textbf{Question}: What are the Banzhaf indices of the players?\\

\section{Extreme cases}\label{SecClassConst}


If the WVG $v$ is $[q;\underbrace{u,u,\ldots, u}_{n}]$, then the Banzhaf indices $\beta_1$, ... , $\beta_n$ are equal to $1/n$. The Banzhaf indices can be found in constant time, and  the following theorem gives the actual number of swings for each player.

\begin{theorem}
In a WVG with $n$ equal weights, $u$, each player is critical in ${{n-1} \choose {\ceil{q/u}}-1}$ coalitions. Moreover, the total number of winning coalitions $w$ is $\sum_{i=\ceil{q/u}}^{n} {n \choose i}$.
\end{theorem}

\begin{proof}
The minimum number of players needed to form a winning coalition is $\ceil{q/u}$. A player is critical in a coalition if there are exactly  $\ceil{q/u}-1$ other players in the coalition. There are ${{n-1} \choose {\ceil{q/u}}-1}$ such coalitions. There are ${n \choose i}$ coalitions of size $i$ and such a coalition is winning if $i \geq \ceil{q/u}$.\qed
\end{proof}

Also, in a WVG with $n$ equal weights $u$, the probabilistic Banzhaf index of each player is then ${{{n-1} \choose {\ceil{q/u}}-1}}/2^{n-1}$. We can also compute Coleman's power of the collectivity to act, $A$, which is equal to $\frac{w}{2^n}$.

A \emph{dictator} is a player who is present in every winning coalition and absent from every losing coalition. This means that the player $1$ with the biggest weight is a dictator if and only if $w_1 \geq q$ and $\sum_{2 \leq i \leq n } w_i<q $. In that case, $\beta_1=1$ and $\beta_i=0$ 
for all $i>1$.


If $0< q \leq w_n$ then the only minimal winning coalitions are all the singleton coalitions.  So there are $n$ minimal winning coalitions and every player is critical in one coalition. Thus, for all $i$, $\beta_i=1/n$ and the Banzhaf indices can be found in constant time (i.e., $\mathcal{O}(1)$). Moreover, the probablistic Banzhaf index $\beta_i^{'}= 1/{2^{n-1}}$ for all $i$, and Coleman's power of collectivity to act $A=\frac{2^n-1}{2^n}$


If $q \geq \sum_{1 \leq i \leq n}w_i - w_n$, then the only minimal winning coalition is $\{1,2,\ldots, n\}$ and it becomes losing if any player gets out of the coalition. Thus the weighted voting game acts like the unanimity game. Then for all $i$, $ \beta_i=1/n$. The Banzhaf indices can be found in constant time (i.e., $\mathcal{O}(1)$). Moreover, for all $i$, $\beta_i^{'}= 1/{2^{n-1}}$ and $ A=1/{2^n}$.

\section{Bounded number of weight values}\label{SecClassBounded}

In this section we estimate the time complexity of several algorithms. We start off with the case when all weights except one are equal and give exact formulas for the Banzhaf indices. We then use this as a warm up exercise to consider more general cases where there are $2$ weight values and then $k$ weight values.

\subsection{All weights except one are equal}

We start off with the case when all weights except one are equal.

\begin{theorem}
Let $v$ be a WVG, $[q;w_a, w_b,...,w_b]$, where there is $w_a$ and $m$ weights of value 
$w_b$, where $w_b < q$. Let $x$ be $\ceil{\frac{q-w_a}{w_b}}$ and  
$y=\ceil{q/w_b}$. Then the total number of coalitions in which a player with weight $w_b$ is critical is ${{m-1}\choose {y-1}}+{{m-1} \choose x}$. Moreover, the number of coalitions in which the player with weight $w_a$ is critical is $\sum_{i=x}^{\mathsf{Min}(y-1,m)} {m \choose i}$. 
\label{thm:easy1} 
\end{theorem}

\begin{proof}
A player with weight $w_b$ is critical in 2 cases:
\begin{enumerate}
\item \textit{It makes a winning coalition with other players with weight $w_b$ only.} Let $y$ be the minimum number of players with weight $w_b$ which form a winning coalition by themselves. Thus $y=\ceil{q/w_b}$. The number of such coalitions in which a player with weight $w_b$ can be critical is ${{m-1}\choose {y-1}}$.  
\item \textit{It makes a winning coalition with the player with weight $w_a$ and none or some players with weight $w_b$}. Let $x$ be the minimum number of players with weight $w_b$ which can form a winning coalition with the inclusion of the player with weight $w_a$. Thus $x= \ceil{\frac{q-w_a}{w_b}}$. Then, the number of such coalitions in which a player with weight $w_b$ can be critical is ${{m-1} \choose x}$. 
\end{enumerate}
The total number of swings for a player with weight $w_b$ is thus ${{m-1}\choose {y-1}}+{{m-1}\choose {x}}$.

The player with weight $w_a$ is critical if it forms a winning a coalition with some players with weight $w_b$ but the coalition becomes losing with its exclusion. The player with weight $w_a$ can prove critical in coalition with varying number of players with weight $w_b$. The maximum number of players with weight $w_b$ with which it forms a winning coalition and is also critical is $y-1$ in case $y\leq m$ and $m$ in case $y > m$. Therefore the total number of coalitions in which the player with weight $w_a$ is critical is $\sum_{i=x}^{\mathsf{Min}(y-1,m)} {m \choose i}$.\qed
\end{proof}

\subsection{Only two different weight values}

Unlike Theorem~\ref{thm:easy1}, we do not give a short formula for the Banzhaf values in the next theorem. However Theorem~\ref{thm:easy2} considers a more general case than Theorem~\ref{thm:easy1}. As we shall we later Theorem~\ref{thm:easy1} provides us with an idea to consider the case of $k$ weight values.  

\begin{theorem}
For a WVG with $n$ players and only two weight values, the Banzhaf indices and numbers of swings can be computed in $\mathcal{O}(n^2)$ time.
\label{thm:easy2}
\end{theorem}

\begin{proof}
We look at a WVG, $v=[q;w_a,...w_a, w_b,...w_b]$, where there are $n_a$ players with weight $w_a$ and $n_b$ players with weight $w_b$. We analyse the situation when a player with weight $w_a$ proves to be critical in a coalition which has $i$ other players with weight $w_a$ and the rest with weight $w_b$. Then the minimum number of players with weight $w_b$ required is $\ceil{\frac{q-(i+1)w_a}{w_b}}$. Moreover the maximum number of players with $w_b$ is $\ceil{\frac{q-iw_a}{w_b}}-1$. Therefore $j$, the number of players with weight $w_b$, satisfies the following inequality: 
$x_1(i)= \ceil{\frac{q-(i+1)w_a}{w_b}} \leq j \leq 
\mathsf{Min}(\ceil{\frac{q-iw_a}{w_b}}-1, n_b)=x_2(i).$ 
Let $A_i = {{n_a-1}\choose i}$, and let $B_i =
\sum_{j=x_1(i)}^{x_2(i)}{n_b \choose j}$. We define, the maximum possible number of extra players with weight $a$, to be $\mathsf{maxa} = \mathsf{Min}(\ceil{q/w_a}-1, n_a-1)$. Then the total number of swings of the player with weight $w_a$ is $\sum_{i=0}^{\mathsf{maxa}} A_i{B_i}$. The total number of swings for a player with weight $w_b$ can be computed by a symmetric method.\qed
\end{proof}

We can devise an algorithm (Algorithm~\ref{BIsFor2ValueWVG}) from the method outlined in the proof.

\begin{algorithm}[H]
  \caption{SwingsFor2ValueWVG}
  \label{SwingsFor2ValueWVG}
  \textbf{Input:} $v=[q;(n_a,w_a),(n_b,w_b)]$.

\textbf{Output:} Total swings of a player with weight $w_a$.

   \begin{algorithmic}[1]

 \STATE $\mathsf{swings}_a \leftarrow 0$\\
 \STATE $\mathsf{maxa} \leftarrow \mathsf{Min}(\ceil{q/w_a}-1, n_a-1)$ 

 \FOR{$i=0$ to $\mathsf{maxa}$}
 
 \STATE $x_1(i)\leftarrow \ceil{\frac{q-(i+1)w_a}{w_b}}$
 \STATE $x_2(i) \leftarrow \mathsf{Min}(\ceil{\frac{q-i(w_a)}{w_b}}-1, n_b)$
 \STATE $A_i \leftarrow {{n_a-1} \choose i}$

  \IF{$x_1(i)>n_b$}
  \STATE $B_i \leftarrow 0$
  \ELSIF{$x_2(i)<0$} 
  \STATE $B_i \leftarrow 0$
  \ELSE
  
  \STATE $B_i \leftarrow 0$
  
 \FOR{$j=x_1(i)$ to $x_2(i)$}
 \STATE $B_i \leftarrow B_i + {n_b \choose j}$
 \ENDFOR
  
  \ENDIF 
\STATE $\mathsf{swings}_a = \mathsf{swings}_a + A_i{B_i}$ 
 \ENDFOR
  
  \RETURN $\mathsf{swings}_a$
  
  \end{algorithmic}
\end{algorithm}

\begin{algorithm}[H]
  \caption{BIsFor2ValueWVG}
  \label{BIsFor2ValueWVG}
  \textbf{Input:} $v=[q;(n_a,w_a),(n_b,w_b)]$.

\textbf{Output:} Banzhaf indices, $\beta = (\beta_a, \beta_b)$.

   \begin{algorithmic}[1]
 
   \STATE $\mathsf{swings}_a=\mathsf{SwingsFor2ValueWVG}(v)$
   \STATE $v'=[q;(n_b,w_b),(n_a,w_a)]$
   \STATE $\mathsf{swings}_b=\mathsf{SwingsFor2ValueWVG}(v')$
   \STATE $\mathsf{totalswings}= n_a\mathsf{swings}_a+n_b\mathsf{swings}_b$
   \STATE $\beta_a = \frac{\mathsf{swings}_a}{\mathsf{totalswings}} $
   \STATE $\beta_b = \frac{\mathsf{swings}_b}{\mathsf{totalswings}} $
   
    \RETURN $(\beta_a, \beta_b)$
  \end{algorithmic}
\end{algorithm}

The algorithm for 2 weight values serves as warm-up for the general case of $k$ weight values in the next section.

\subsection{$k$ weight values}

\begin{theorem}
The problem of computing Banzhaf indices of a WVG with $k$ possible values of the weights is solvable in $O(n^{k})$.
\end{theorem}
\begin{proof}
We can represent a WVG $v$ with $k$ weight classes as follows: 
$$[q;(n_1, w_1),(n_2,w_1),...,(n_k,w_k)]$$ where $n_{i}$ is the number of players with weights $w_{i}$ for $i=1, \ldots, k$. Here, we extend Algorithm~\ref{BIsFor2ValueWVG} to Algorithm~\ref{BIsFor-k-ValueWVG} for $k$ weight classes.

We can write $v'$ as $[q;(1,w_0), (n_1-1, w_1),...,(n_k,w_k)]$ where $w_0=w_1$. This makes it simpler to write a recursive function to compute the number of swings of player with weight $w_0$.
Let $A_{i_1,i_2,...,i_{m}}$ be the number of swings for $w_0$ where there are $i_j$ players with weight $w_j$ in the coalition for $1 \leq j \leq m$.

Then
$$A_{i_1,i_2,...,i_k} = \left \{ \begin{array}{ll}
{n_1-1 \choose i_1}(\Pi_{j=2}^k {n_j \choose i_j}) &
\textrm{if $q-w_0\leq \sum_{j=1}^k{i_j}w_{j} < q$} \\
0 &  \textrm{otherwise.}\\
\end{array} \right.$$

Now for $1\leq m\leq k$,
$$A_{i_1,i_2,...,i_{m-1}} =  \sum_{i_m} A_{i_1,i_2,...,i_{m}}.$$

Here the summation is taken over all values of $i_m$ for which the contribution is non-zero. Explicitly, this range is given by
$$\mathsf{Max}(\left\lceil \frac{q - w_{0} - \sum_{j=1}^{m-1}{i_j}w_{j} - \sum_{j=m+1}^{k}{n_j}w_{j}}{w_{m}}\right\rceil ,0)
\leq i_m \leq
\mathsf{Min}(\left\lceil\frac{q - \sum_{j=1}^{m-1}{i_j}w_{j}}{w_{m}}\right\rceil-1, n_{m}) .$$

The total number of swings of the player with weight $w_0$ is then $A_{\epsilon}$.

\end{proof}

\begin{algorithm}[H]
  \caption{SwingsForWVG}
  \label{SwingsFor-k-ValueWVG}
  \textbf{Input:} $v=[q;(n_1,w_1),(n_1,w_1),\ldots, (n_k,w_k)]$.

\textbf{Output:} Total number of swings, $\mathsf{swings}_0$, of a player with weight $w_1$.

  \begin{algorithmic}[1]
 
  \STATE $w_0=w_1$
  \STATE $v'=[q;(1,w_0), (n-1, w_1),...,(n_k,w_k)]$
  \STATE $\mathsf{swings}_0= A_{\epsilon}$
    \RETURN $\mathsf{swings}_0$
  
  \end{algorithmic}
\end{algorithm}

\begin{algorithm}[H]
  \caption{BIsFor-$k$-ValueWVG}
  \label{BIsFor-k-ValueWVG}
  \textbf{Input:} $v=[q;(n_1,w_1),(n_1,w_1),\ldots, (n_k,w_k)]$.

\textbf{Output:} Banzhaf indices, $\beta = (\beta_1,\ldots \beta_k)$.

   \begin{algorithmic}[1]
 
   \STATE $\mathsf{swings}_1=\mathsf{SwingsForWVG}(v)$
   \STATE $\mathsf{totalswings} \leftarrow 0 $   
   \FOR{$i=2$ to $k$}
   \STATE $v=\mathsf{Swap}(v,(n_1,w_1)(n_i, w_i))$
   \STATE $\mathsf{swings}_i= \mathsf{SwingsForWVG}(v)$
   \STATE $\mathsf{totalswings} \leftarrow \mathsf{totalswings}+ n_i{\mathsf{swings}_i}$
   \ENDFOR
   
    \FOR{$i=1$ to $k$}
   \STATE $\beta_i=\frac{\mathsf{swings}_i}{\mathsf{totalswings}}$
   \ENDFOR

    \RETURN $(\beta_1,\ldots \beta_k)$
   
  \end{algorithmic}
\end{algorithm}

We note that the exact computational complexity of BI-WVG for a WVG with $k$ weight values is $\mathcal{O}(k{(\frac{n}{k})}^k)$ where ${(\frac{n}{k})}^k\geq n_1\cdots n_k$. 
None of the algorithms presented for WVGs with bounded weight values extends naturally for multiple weighted voting games.

\section{Distribution of weights}\label{SecClassDist}

\subsection{Geometric sequence of weights, and unbalanced weights}

\begin{definition}
An $r$-\textit{geometric} WVG $[q;w_1,... , w_n]$ is a WVG where $w_i \geq r{w_{i+1}} $ for $i=1,...,n-1$.
\end{definition}

We observe that in a 2-geometric WVG (such as $[q;2^n, 2^{n-1},...,]$), for any target sum of a coalition, we can use a greedy approach, trying to put bigger weights first, to come as close to the target as possible.  This greedy approach was first identified by Chakravarty, Goel and Sastry~\cite{RePEc:eee:matsoc:v:40:y:2000:i:2:p:227-235} for a broader category of weighted voting games in which weights are \textit{unbalanced}:

\index{unbalanced WVG}
\begin{definition}
An \emph{unbalanced WVG} is a WVG such that, for $1 \leq j \leq n$, $w_j>w_{j+1}+w_{j+2}...+w_n$. 
\end{definition}

\begin{example}
The game $[22;18,9,4,2,1]$ is an example of an unbalanced WVG where each weight is greater than the sum of the subsequent weights.
\end{example}

Chakravarty, Goel and Sastry \cite{RePEc:eee:matsoc:v:40:y:2000:i:2:p:227-235} showed that the greedy approach for unbalanced WVG with integer weights can help to compute all Banzhaf indices in $\mathcal{O}(n)$. We notice that the same algorithm can be used for an unbalanced WVG with real weights without any modification. In fact it is this property of `geometric weights' being unbalanced which is the reason that we can find suitable coalitions for target sums so efficiently. We characterise those geometric sequences which give unbalanced WVGs: 

\begin{theorem}
If $r\geq 2$ then every $r$-geometric WVG is unbalanced.  
\end{theorem}

\begin{proof}
Let $v$ be an $r$-geometric WVG. We prove by induction that $w_j>w_{j+1}+\ldots+w_n$. This is true for $j=n$. Suppose it is true for all $i$, $j+1\leq i\leq n$. Since $v$ is $r$-geometric, $w_j\geq 2w_{j+1}$. But, $2w_{j+1}=w_{j+1}+w_{j+1}> w_{j+1}+w_{j+2}+\ldots+w_n$. Therefore $v$ is unbalanced. \qed
\end{proof}

\begin{corollary}
For an $r$-geometric WVG $v$ where $r \geq 2$, the Banzhaf indices of players in $v$ can be computed in $\mathcal{O}(n)$ time. 
 
\end{corollary}

\begin{proof}
Since the condition of $r\geq 2$ makes $v$ an unbalanced WVG, then we can use the greedy algorithm from \cite{RePEc:eee:matsoc:v:40:y:2000:i:2:p:227-235} which computes the Banzhaf indices in $\mathcal{O}(n)$. \qed
\end{proof}


\index{$k$-unbalanced WVG}
\begin{definition}
A WVG is a \emph{$k$-unbalanced WVG} if, for $1 \leq j \leq n$, $w_j>w_{j+k}+\cdots +w_n$. So an unbalanced WVG is `1-unbalanced'.
\end{definition}

Note that an $r$-geometric WVG is 2-unbalanced when 
$r \geq \frac{1 + \sqrt{5}}{2} \approx 1.61803... = \varphi$, 
the golden ratio, since then
$$\frac{1}{r^2}+\frac{1}{r^3}+\cdots < \frac{1}{r(r-1)}
\leq 1\ \mbox{\rm since\ } r(r-1)\geq \varphi(\varphi-1) = 1.$$

We check whether 2-unbalanced WVGs have properties similar to those of unbalanced WVGs. 

\begin{example}\label{3game}
Consider a WVG $v$ with $2m$ players and weights 
$$3^{m-1},3^{m-1},\ldots,3^j,3^j,\ldots,3,3,1,1.$$ 
It is easy to see that $\sum_{i=0}^{j-1} 2\cdot 3^i < 3^j$, so the game is 2-unbalanced. 
\end{example}

In the unbalanced game, for each target coalition sum, there is either one corresponding coalition or none. This does not hold for 2-unbalanced WVGs. 
In Example~\ref{3game} with target total $1+3+\cdots+3^{m-1} = \frac12(3^m-1)$, 
there are exactly $2^m$ coalitions which give this target, namely those 
coalitions with exactly one player out of each equal pair.

We prove that even for the class of 2-unbalanced (instead of simply unbalanced WVGs) 
the problem of computing Banzhaf indices becomes NP-hard. 
\begin{theorem}
BI-WVG is NP-hard for the class of 2-unbalanced WVGs .
\label{NP-hard-2unbalanced}
\end{theorem}

\begin{proof} 
We will use a reduction from the following NP-hard problem:\\ 

\noindent
\textbf{Name}: \textit{SUBSET SUM}   \\
\textbf{Instance}: $z_1, \ldots ,z_m$, $T \in \mathrm{N}$. \\
\textbf{Question}: Are there $x_j$s in $\{0,1\}$ so that $\sum_{j=1}^m x_j z_j= T$?\\

For the reduction from SUBSET SUM, we scale and modify the weights from the WVG $v$ of
Example~\ref{3game}. 
For any instance $I=\{z_1, \ldots ,z_m, T\}$ of SUBSET SUM, we will define a game $v_I$ 
with $2m+1$ players. 
Let $Z=1+\sum_{j=1}^m z_j$, and we may assume that $T<Z$. 
Whereas $v$ had pairs of weights $3^j,3^j$ for $0\leq j \leq m-1$, 
in $v_I$ there is one ``unit player'' with weight~1 and $2m$ pairs of players 
with weights $3^j Z, 3^j Z +z_j$ for $0\leq j \leq m-1$. The quota for $v_I$ 
is $\frac12 (3^m-1)Z + T + 1$. The unit player has nonzero Banzhaf index 
if and only if there exists a coalition among the other $2m$ players with weight 
exactly $\frac12 (3^m-1)Z + T$. We will show that to determine this is 
equivalent to answering the SUBSET SUM instance $I$, and so even this 
special case of BI-WVG is NP-hard.

In Example~\ref{3game}, it was necessary (and sufficient) for achieving the 
target total of $\frac12(3^m-1)$ to take exactly one player from each pair. 
In game $v_I$, since $\sum_{j=1}^m z_j < Z$, this is still a necessary condition 
for achieving the total of $\frac12 (3^m-1)Z + T$, and whether or not there 
is such a selection achieving the total is exactly the condition of whether there 
is a subset of the $z_j$s which sums to~$T$.\qed
\end{proof}

\subsection{Sequential weights}

\begin{definition}
The set of weights $\{w_1,w_2,... ,w_n\}$ is \emph{sequential} if 
$$w_n|w_{n-1}|w_{n-2}...|w_1,$$ 
\noindent
i.e. each weight is a multiple of the next weight.
\end{definition}

\begin{example}
$[32;20, 10, 10, 5, 1, 1, 1]$ is an example of a WVG with sequential weights.
\end{example}

Chakravarty, Goel and Sastry \cite{RePEc:eee:matsoc:v:40:y:2000:i:2:p:227-235} show that Banzhaf indices can be computed in $\mathcal{O}(n^2)$ time if the weights are sequential and they satisfy an additional dominance condition. The diminance conditions states that a weight in one weight class should be more than the sum of weights of any subsequent weight class. 

\index{dominance condition}

\begin{definition}
Let $d_1>d_2>\cdots >d_r$ be the distinct values of weights $w_1,\ldots ,w_n$ of a sequential set. Then $d_k=m_k{d_{k+1}}$ where $m_k>1$, $\forall k$, $1 \leq k < r$. Let $N_k=\{i~~|~~w_i=d_k\}$ and $n_k=|N_k|$. Then the \emph{dominance condition} holds if $m_k>n_{k+1}$ $\forall k$, $1 \leq k < r$.
\end{definition} 

\index{alternative dominance condition}

We now define the alternative dominance condition for WVGs.
\begin{definition}
Let $d_1>d_2>\cdots >d_r$ be the distinct values of weights $w_1,\ldots ,w_n$ of a sequential set. Let $N_k=\{i|w_i=d_k\}$ and $n_k=|N_k|$. Then the \emph{alternative dominance condition} holds if $\forall j \in N_k$, $1\leq k<r$, $w_j>\sum \{w_p~|~p\in N_i, i>k \}$.
\end{definition} 


We provide an alternative dominance condition for weights which are not necessarily sequential. It is easy to see that a 2-unbalanced WVG does not necessarily satisfy the alternative dominance condition.

\begin{definition}
Let $d_1>d_2>\cdots >d_r$ be the distinct values of weights $w_1,\ldots ,w_n$ of a sequential set. Let $N_k=\{i|w_i=d_k\}$ and $n_k=|N_k|$. Then the \emph{alternative dominance condition} holds if $\forall j \in N_k$, $1\leq k<r$, $w_j>\sum \{w_p~|~p\in N_i, i>k \}$.
\end{definition} 

\begin{proposition}
Suppose a WVG $v$ satisfies the alternative dominance condition. Then for $v$, BI-WVG has time complexity $\mathcal{O}(n^2)$.
\end{proposition}

\begin{proof}
This follows from Theorem 10 in \cite{RePEc:eee:matsoc:v:40:y:2000:i:2:p:227-235} where the proof is for a sequential WVG which obeys the dominance condition. However we notice that since the argument in the proof can be made for any WVG which satisfies the alternative dominance condition, the proposition holds for $v$.  \qed
\end{proof}

\section{Integer weights}\label{SecClassInteger}

When all weights are integers, other methods may become applicable.

\subsection{Moderate sized integer weights}
Matsui and Matsui~\cite{matsui00survey} prove that a dynamic programming approach provides a pseudo-polynomial algorithm to compute Banzhaf indices of all players with time complexity $\mathcal{O}({n^2}q)$. Since $q$ is less than $\sum_{i \in N}w_i$, the Banzhaf indices can be computed in polynomial time if the weight sizes are moderate.

\subsection{Polynomial number of coefficients in the generating function of the WVG}

A generating function is a formal power series whose coefficients encode information about a sequence.
Bilbao et al.~\cite{bilbao_generating} observe, for a WVG $v=[q;w_1,\ldots ,w_n]$, that if the number of coalitions for which a player $i$ is critical is 
$b_i = |\{S \subset N : v(S)=0, v(S \cup \{i\})=1 \}| \allowbreak
= \sum^{q-1}_{k=q-w_i} b^{i}_{k} $, where $b^{i}_{k}$ is the number of coalitions 
which do not include $i$ and with total weight $k$, then  the generating functions 
of the numbers $\{b^{i}_{k} \}$ are given by 
$B_i(x)= {\prod}^{n}_{j=1, j\neq i} (1+x^{w_j})
= 1+ {b^{i}_{1}}x +{b^{i}_{2}}x^2+\cdots +{b^{i}_{W-w_i}}x^{W-w_i}$. 
This was first pointed out by Brams and Affuso~\cite{brams}. 

\begin{example}
Let $v=[6;5,4,1]$ be a WVG.

\begin{itemize}
\item ${B_1}(x)= (1+x^4)(1+x^1)= 1 + x + x^4 + x^5$

The coalitions in which player $1$ is critical are $\{1,2\}$, $\{1,3\}$, $\{1,2,3\}$. Therefore ${\eta}_1= 3$.
\item ${B_2}(x)= (1+x^5)(1+x^1)= 1 + x + x^5 + x^6$
The coalition in which player $2$ is critical is $\{1,2\}$. Therefore ${\eta}_2= 1$.
\item ${B_3}(x)= (1+x^5)(1+x^4)= 1 + x^4 + x^5 + x^9$

The coalition in which player $3$ is critical is $\{1,3\}$. Therefore ${\eta}_3=1$.
\end{itemize}
Consequently, $\beta_1 = 3/5$, $\beta_2 = 1/5$ and $\beta_3 = 1/5$.
\end{example}

The generating function method provides an efficient way of computing Banzhaf indices if the voting weights are moderate integers. 
Bilbao et al. \cite{bilbao_generating} prove that the computational complexity of computing Banzhaf indices by generating functions is $\mathcal{O}(n^2{C})$ where $C$ is the number of non-zero coefficients in $\prod_{1\leq i \leq n}(1+x^{w_j})$. We note that $C$ can be bounded by the sum of the weights but the bound is not tight. $C$ can be relatively small even if the weight values are exponential in $n$. Therefore if a $WVG$ has a generating function in which the number of non-zero terms is polynomial in $n$, then the computational complexity of computing the Banzhaf indices is in $P$.

\section{Open problems \& conclusion}\label{SecClassOpen}

Table~\ref{easywvg-classes} contains a summary of the algorithms or complexity results for different classes of WVGs. A\&P refers to Aziz and Paterson. In this paper we have classified WVGs for which Banzhaf indices can be computed in polynomial time. It would be interesting to identify further important classes of WVGs which have less than exponential time complexity. The extensive literature on the SUBSET-SUM problem should offer guidance here. It appears an interesting question to analyse the expected number of terms in the generating function for sequential WVGs. Another challenging open problem is to devise an algorithm to compute exactly the Banzhaf indices of a general WVG in time complexity which is less than $\mathcal{O}(n^2 2^{\frac{n}{2}})$.

\begin{table}[h!b!p!]
\small
\caption{Complexity of WVG classes}
\begin{center}
\begin{tabular}{llllr}
\toprule
	
\textbf{WVG Class}& \textrm{R}/\textrm{Z}~~~~&Complexity&Time&Remarks\\
& &Class&&\\
\midrule

General&R/Z&NP-Hard&$\mathcal{O}(n^2 1.415^n)$&\cite{RePEc:eee:matsoc:v:49:y:2005:i:1:p:111-116}\\

Unbalanced&R/Z&P&$\mathcal{O}(n)$&\cite{RePEc:eee:matsoc:v:40:y:2000:i:2:p:227-235}\\

k-Unbalanced($k\geq 2$)&R/Z&NP-Hard&&A\&P~\cite{Aziz-classification}\\

Sequential with dominance& R/Z & P & $\mathcal{O}(n^2)$&\cite{RePEc:eee:matsoc:v:40:y:2000:i:2:p:227-235}\\

Alternative dominance& R/Z&P& $\mathcal{O}(n)$&A\&P~\cite{Aziz-classification} + \cite{RePEc:eee:matsoc:v:40:y:2000:i:2:p:227-235}\\

Bounded(k)~\#(weight values)& R/Z&P& $\mathcal{O}(n^k)$&A\&P~\cite{Aziz-classification}\\

$r$-geometric&R/Z&P&$\mathcal{O}(n)$&A\&P~\cite{Aziz-classification} + \cite{RePEc:eee:matsoc:v:40:y:2000:i:2:p:227-235}\\

Moderate integer weights&Z&P& $\mathcal{O}({n^2}q)$&\cite{matsui00survey}\\

Moderate GF&Z&P&$\mathcal{O}({n^2}C)$&\cite{brams}, \cite{bilbao_generating}\\

\bottomrule

\end{tabular}
\end{center}
\normalsize
\label{easywvg-classes}
\end{table}

\bibliographystyle{plain}


{\bf HARIS AZIZ}
\\
{\it Department of Computer Science, University of Warwick, 
Coventry CV4 7AL, United Kingdom.\\
haris.aziz@warwick.ac.uk.}

{\bf MIKE PATERSON}
\\
{\it Department of Computer Science, University of Warwick, 
Coventry CV4 7AL, United Kingdom.\\
msp@dcs.warwick.ac.uk.}

\end{document}